\documentclass[twocolumn,showpacs,preprintnumbers,amsmath,amssymb,superscriptaddress]{revtex4}
\usepackage{color}
\usepackage{epsfig,bm}
\newcommand{\vi}[1]{\mbox{\boldmath $#1$}}

\begin{document}
\title{Ab initio study of the photoabsorption of $^4$He} 
\author{W. Horiuchi}%
\affiliation{RIKEN Nishina Center, Wako 351-0198, Japan}
\author{Y. Suzuki}
\affiliation{Department of Physics, Niigata University, Niigata
950-2181, Japan}
\affiliation{RIKEN Nishina Center, Wako 351-0198, Japan}
\author{K. Arai}
\affiliation{Division of General Education, Nagaoka National College of
Technology, Nagaoka 940-8532, Japan}

\begin{abstract}
There are some discrepancies in the low energy 
data on the photoabsorption cross section of $^4$He.  We calculate  
the cross section with realistic nuclear forces and 
explicitly correlated Gaussian functions. 
Final state interactions and two- and three-body decay 
channels are taken into account. The cross section is 
evaluated in two methods: With the complex scaling method 
the total absorption cross section is obtained 
up to the rest energy of a pion, 
and with the microscopic $R$-matrix method both cross sections 
$^4$He($\gamma, p$)$^3$H and $^4$He($\gamma, n$)$^3$He are 
calculated below 40\,MeV. 
Both methods give virtually the same result. 
The cross section rises sharply from the 
$^3$H+$p$ threshold, reaching a giant resonance peak at 26--27\,MeV. 
Our calculation reproduces almost all the data above 30\,MeV. 
We stress the importance of 
$^3$H+$p$ and $^3$He+$n$ cluster configurations on the cross 
section as well as the effect of 
the one-pion exchange potential on 
the photonuclear sum rule. 
\end{abstract}
\maketitle

\section{Introduction}

Nuclear  strength or response functions for electroweak 
interactions provide us with important information 
on the resonant and continuum structure of the nuclear system 
as well as the detailed property 
of the underlying interactions. In this paper we focus on 
the photoabsorption of $^4$He. The experimental 
study of $(\gamma, p)$ and $(\gamma, n)$ reactions on $^4$He has 
a long history over the last half century. 
See Refs.~\cite{shima,nilsson,quaglioni04} and 
references therein. Unfortunately  the experimental  
data presented so far are in serious disagreement, and thus 
a measurement of the photoabsorption cross section is still 
actively performed with different techniques 
in order to resolve this enigma~\cite{nakayama,tornow}. 

Calculations of the cross section on $^4$He  
have been performed in several methods focusing on e.g., the 
peak position of the giant electric dipole ($E1$) resonance,  
charge symmetry breaking effects, and $E1$ sum 
rules~\cite{efros,wachter,gazitb}.  
The photoabsorption cross section  has 
extensively been calculated in  the Lorentz integral transform 
(LIT) method~\cite{LIT}, among others, which does not require calculating  
continuum wave functions. In the 
LIT the cross section is obtained by inverting the  
integral transform of the strength function, which 
is calculable using square-integrable (${\cal L}^2$) functions.  
The calculations were done with Malfliet-Tjon central 
force~\cite{quaglioni04}, the realistic Argonne $v$18  
potential~\cite{gazit,bacca}, and an interaction 
based on chiral effective field theory~\cite{quaglioni}. 

In the calculations with the realistic interactions 
some singular nature of them,  
especially the short-range repulsion, has been 
appropriately replaced with the effective one 
that adapts to the model space of the respective approaches, that is, 
the hyperspherical harmonics method~\cite{gazit,bacca}
and the no-core shell model~\cite{quaglioni}. All of these calculations 
show the cross section that disagrees with the data~\cite{shima} 
especially in the low excitation energy 
near the $^3$H+$p$ threshold. The resonance peak obtained 
theoretically appears at about 27 MeV consistently 
with the experiments~\cite{nilsson,nakayama,tornow}, but in a 
marked difference from that of 
Ref.~\cite{shima}.

We have recently reported that all the observed
levels of $^4$He below 26 MeV are well reproduced 
in a four-body calculation using bare
realistic nuclear interactions~\cite{dgvr,inversion}.
It is found that using the realistic interaction is vital 
to reproduce the $^4$He spectrum as well as the 
well-developed $3N$+$N$ ($^3$H+$p$ and $^3$He+$n$) 
cluster states with positive and negative parities. In this 
calculation the wave functions of the states are approximated 
in a combination of explicitly correlated 
Gaussians~\cite{boys,singer} reinforced with a global vector  
representation for the angular motion~\cite{varga,svm}. 
Furthermore this approach has very recently  been  
applied to successfully describe four-nucleon 
scattering and reactions~\cite{arai11,fbaoyama} with the aid of 
a microscopic
$R$-matrix method (MRM)~\cite{desc}. It is 
found that the tensor force plays a crucial role in accounting for 
the astrophysical $S$ factors of the radiative 
capture reaction $^2$H$(d,\gamma)^4$He as well as the nucleon 
transfer reactions, $^2$H$(d,p)^3$H and $^2$H$(d,n)^3$He~\cite{arai11}.

The aim of this paper is to examine the issue of the 
photoabsorption cross section of $^{4}$He. Because 
four-body bound state problems with realistic nucleon-nucleon ($NN$) 
interactions can be accurately solved with the correlated Gaussians,  
it is interesting to apply that approach 
to a calculation of the photoabsorption cross section. 
For this purpose we have to convert the continuum problem to 
such a bound-state like problem that can be treated in the 
${\cal L}^2$ basis functions. 
Differently from the previous theoretical 
calculations~\cite{quaglioni04,gazit,bacca,quaglioni}, 
we employ a complex scaling method (CSM)~\cite{ho,moiseyev,CSM} 
for avoiding a construction of the continuum wave functions. 
One of the advantages of the CSM is that the cross section can be 
directly obtained without recourse to a sophisticated inversion 
technique as used in the LIT or an artificial energy averaging 
procedure.  We will pay 
special attention to the following points:
\begin{enumerate}
\vspace{-1mm}
\item To use a realistic interaction as it is 
\vspace{-2mm}
\item To include couplings with final decay channels explicitly 
\vspace{-2mm}
\item To perform calculations in both MRM and CSM as a cross-check.
\vspace{-1mm}
\end{enumerate}
Here point (1) indicates that the interaction is not changed to 
an effective force by some transformation.
This looks sound and appealing because 
the cross section may depend on the $D$-state 
probability of $^4$He~\cite{wachter} and hence the effect of the tensor force 
on the cross section could be seen directly. 
In point (2) we make use of the flexibility of the correlated Gaussians  
to include such important configurations that have 
$^3$H+$p$, $^3$He+$n$, and $d$+$p$+$n$ partitions. Thanks to this 
treatment the effects of final-state interactions are expected to be 
fully taken into account. 
Point (3) is probably most significant in our approach. 
We mean by this point that the photoabsorption cross section 
is calculated in two independent methods. In the MRM we calculate the 
cross sections for the radiative capture reactions, 
$^3$H$(p,\gamma)^4$He and 
$^3$He$(n,\gamma)^4$He, and these cross sections are converted to 
the photoabsorption cross section using a formula due to 
the detailed balance. In the CSM we 
make use of the fact that the final 
continuum states of $^4$He, if rotated on the complex coordinate 
plane, can be expanded in the ${\cal L}^2$ functions. Consistency 
of the two results, if attained, serves strong evidence 
for that the obtained cross section is reliable. 
We hope to shed light on resolving the controversy
from our theoretical input. 

In Sec.~\ref{continuum.sec} 
we present our theoretical prescriptions
to calculate the photoabsorption cross section. 
The two approaches, the CSM and the MRM, are explained in this section 
with emphasis on the method of how discretized states are employed 
for the continuum problem. We give the basic inputs of our calculation 
in Sec.~\ref{model.sec}. The detail of our correlated 
basis functions 
is given in Sec.~\ref{basisfunction}, and various configurations needed to 
take into account the final state interactions as well as two- and 
three-body decay channels are explained in 
Sec.~\ref{squareintegrable}. We show 
results on the photoabsorption cross section in Sec.~\ref{results.sec}. 
The $E1$ strength function and the transition densities  
calculated from the continuum discretized states 
are presented in Sec.~\ref{eds.transition}. 
A comparison of CSM and MRM cross sections is made in 
Sec.~\ref{test.CSM-MRM}. The photonuclear sum rules are 
examined in Sec.~\ref{E1.sum.rule}. The calculated photoabsorption 
cross sections are compared to experiment in Sec.~\ref{comparison}. 
Finally we draw conclusions of this work in Sec.~\ref{conclusion.sec}.

\section{Formulation of photoabsorption cross section calculation}
\label{continuum.sec}

\subsection{Basic formula}
The photoabsorption takes place mainly through the 
$E1$ transition, which can be treated by the 
perturbation theory. The wavelength of the photon energy $E_{\gamma}$ 
(MeV) is about $1240/E_{\gamma}$ (fm), so that it is long enough 
compared to 
the radius of $^4$He even when $E_{\gamma}$ is close to the rest 
energy of a pion. 
The photoabsorption cross section $\sigma_\gamma (E_{\gamma})$ can be
calculated by the formula~\cite{ring}
\begin{align}
\sigma_\gamma (E_{\gamma})=\frac{4\pi^2}{\hbar c}E_{\gamma}\frac{1}{3}S(E_{\gamma}),
\label{photo-abs.eq}
\end{align}
where $S(E)$ is the strength function for the $E1$ transition 
\begin{align}
S(E)={\vi {\cal S}}_{\mu f}
|\left<\Psi_f\right|\mathcal{M}_{1\mu}\left|\Psi_0\right>|^2
\delta(E_f-E_0-E).
\label{def.S}
\end{align}
The symbol $\mathcal{M}_{1\mu}$ 
denotes the $E1$ operator, and $\Psi_0$ and 
$\Psi_f$ are the wave functions of the ground state with energy $E_0$ and 
the final state with the excitation energy $E_f$ of $^4$He, respectively. 
The recoil energy of $^4$He is ignored, so that $E_{\gamma}$ is equal to 
the nuclear excitation energy $E$. The symbol ${\vi {\cal S}}_{\mu f}$ 
indicates a summation 
over $\mu$ and all possible final states $f$.
The final state of $^4$He is actually a continuum state lying 
above the $^3$H+$p$ threshold, and it is normalized 
according to $\left<\Psi_{f^\prime}|\Psi_f\right>=
\delta(E_{f^\prime}-E_f)$.  The sum  or integral for 
the final states in 
${\vi {\cal S}}_{\mu f}$ can be taken using the closure 
relation, leading to a well-known expression for 
the strength function 
\begin{align}
S(E)=-\frac{1}{\pi}\text{Im}\sum_{\mu}\left<\Psi_0\right|
\mathcal{M}^\dagger_{1\mu}\frac{1}{E-H+i\epsilon}\mathcal{M}_{1\mu}
\left|\Psi_0\right>,
\label{sf.eq}
\end{align}
where a positive infinitesimal $\epsilon$ ensures the outgoing wave
after the excitation of $^4$He. A method of calculation of 
$S(E)$ in the CSM is presented in Sec.~\ref{csm}.

A partial photoabsorption cross section $\sigma_{\gamma}^{\rm
AB}(E_{\gamma})$ 
for the two-body final state comprising nuclei, A and B, can be 
calculated in another way. 
With use of the detailed balance the cross section is related
to that of its inverse process, the radiative 
capture cross section 
$\sigma_{\rm cap}^{\rm AB}(E_{\rm in})$~\cite{thompson09}, 
induced by the $E1$
transition, at the incident energy $E_{\rm in}=E_{\gamma}-E_{\rm th}$,   
\begin{equation}
\sigma_{\gamma}^{\rm AB}(E_{\gamma}) = \frac{ k^2 (2J_A+1)(2J_B+1)}
{2 k^2_{\gamma} (2J_0+1)} \sigma_{\rm cap}^{\rm AB}(E_{\rm in}),
\end{equation}
where $E_{\rm th}$ is the A+B threshold energy. 
Here $J_A$ and $J_B$ are the angular momenta of the nuclei, $\rm A$ and
$\rm B$,  
and $J_0(=0)$ is the angular momentum of the ground state of
$^4$He. The wave
number $k$ is $\sqrt{2\mu_{AB}E_{\rm in}/\hbar^2}$ where $\mu_{AB}$ is the
reduced mass of the two nuclei and $k_{\gamma}$ is 
the photon wave number $E_{\gamma}/\hbar c$. The 
photoabsorption cross section  
$\sigma_{\gamma}(E_{\gamma})$ is equal to a sum of 
$\sigma_{\gamma}^{^3{\rm H}\,p}(E_{\gamma})$ and
$\sigma_{\gamma}^{^3{\rm He}\,n}(E_{\gamma})$ provided that 
three- and four-body breakup contributions are negligible. A 
calculation of the radiative capture cross section will be performed 
in the MRM as explained in Sec.~\ref{mrm}. 

The fact that we have two independent methods of calculating 
$\sigma_{\gamma} (E_{\gamma})$ is quite important to assess 
their validity.

\subsection{Complex scaling method}
\label{csm}

The quantity $S(E)$ of Eq.~(\ref{sf.eq}) is evaluated using 
the CSM, which makes a continuum state that has an outgoing wave 
in the asymptotic region damp at large distances, thus enabling us 
to avoid an explicit construction of the 
continuum state. In the CSM the single particle coordinate 
$\bm{r}_j$ and momentum $\bm{p}_j$ are subject to a rotation by an angle $\theta$: 
\begin{align}
U(\theta):\ \ \ \ \bm{r}_j \to \bm{r}_j\text{e}^{i\theta},\ \ \ \bm{p}_j
 \to \bm{p}_j\text{e}^{-i\theta}.
\label{trans.eq}
\end{align}
Applying this transformation in  Eq.~(\ref{sf.eq}) leads to 
\begin{align}
S(E)=-\frac{1}{\pi}\text{Im}\sum_\mu\left<\Psi_0\right|
\mathcal{M}_{1\mu}^\dagger U^{-1}(\theta)R(\theta)
U(\theta)\mathcal{M}_{1\mu}\left|\Psi_0\right>,
\end{align}
where $R(\theta)$ is the complex scaled resolvent 
\begin{align}
R(\theta)=\frac{1}{E-H(\theta)+i\epsilon}
\end{align}
with
\begin{align}
H(\theta)=U(\theta)HU^{-1}(\theta).
\end{align}

A key point in the CSM is that within a suitable range of positive 
$\theta$ the eigenvalue problem 
\begin{align}
H(\theta)\Psi_\lambda^{JM\pi}(\theta)=E_\lambda(\theta)\Psi_\lambda^{JM\pi}(\theta)
\label{ceig.eq}
\end{align}
can be solved in a set of $\mathcal{L}^2$ basis functions $\Phi_i (\bm{x})$
\begin{align}
\Psi_\lambda^{JM\pi}(\theta)=\sum_iC_i^\lambda (\theta)\Phi_i(\bm{x}).
\end{align}
We are interested in $\Psi_\lambda^{JM\pi}(\theta)$ with $J^{\pi}=1^-$. 
With the solution of Eq.~(\ref{ceig.eq}), an expression for $S(E)$ reads~\cite{myo,threebody}  
\begin{align}
S(E)=-\frac{1}{\pi}\sum_{\mu \lambda} \text{Im}
\frac{\widetilde{\mathcal{D}}^\lambda_\mu(\theta)\mathcal{D}_\mu^\lambda(\theta)}
{E-E_\lambda(\theta)+i\epsilon},
\label{csm.strength}
\end{align}
where 
\begin{align}
&\mathcal{D}^\lambda_\mu(\theta)=\left<(\Psi_\lambda^{JM\pi}(\theta))^* 
\right|\mathcal{M}_{1\mu}(\theta)\left|U(\theta)\Psi_0 \right>,\notag\\
&\mathcal{\widetilde{D}}_\mu^{\lambda}(\theta)=\left<(U(\theta)\Psi_0 )^* 
\right|{\widetilde{\mathcal{M}}}_{1\mu}(\theta)\left|\Psi_\lambda^{JM\pi}(\theta)\right>,
\end{align}
with
\begin{align}
&\mathcal{M}_{1\mu}(\theta)=U(\theta)\mathcal{M}_{1\mu}U^{-1}(\theta)
=\mathcal{M}_{1\mu}{\rm e}^{i\theta},\notag\\
&\widetilde{\mathcal{M}}_{1\mu}(\theta)=U(\theta){\mathcal{M}}_{1\mu}^{\dagger}
U^{-1}(\theta)=\mathcal{M}_{1\mu}^{\dagger} {\rm e}^{i\theta}.
\end{align}

Note that the energy of the bound state of $H$ in principle remains the same 
against the scaling angle $\theta$. Also 
$U(\theta)\Psi_0$ is to be understood as a solution of
Eq.~(\ref{ceig.eq}) for $J^{\pi}=0^+$ 
corresponding to the ground-state energy~\cite{threebody}. This
stability condition will be met when the 
basis functions are chosen sufficiently. 

In such a
case where sharp resonances exist, the angle $\theta$ has to be rotated to 
cover their resonance poles on the complex energy plane~\cite{moiseyev,CSM}.  
A choice of $\theta$ is made by examining
the stability of $S(E)$ with respect to the angle. 
One of the advantages of the CSM is that one needs no artificial energy
smoothing procedure but obtains the continuous cross section naturally.

\subsection{Microscopic $R$-matrix method}
\label{mrm}

The calculation of $\sigma_{\rm cap}^{\rm AB}(E_{\rm in})$ involves the matrix 
element of ${\cal M}_{1\mu}$ between the scattering state initiated through 
the A+B entrance channel and the final state, i.e., the ground state 
of $^4$He. See, e.g., Ref.~\cite{arai02}. 
The scattering problem is solved in the MRM. 
As is discussed in detail
for the four-nucleon scattering~\cite{arai10,fbaoyama}, an accurate 
solution for the scattering problem with realistic $NN$ potentials 
in general requires a full account 
of couplings of various channels. In the present study 
we include the following two-body channels: $^3$H($\frac{1}{2}^+$)+$p$,
$^3$He($\frac{1}{2}^+$)+$n$, $d$(1$^+$)+$d$(1$^+$), $pn$(0$^+$)+$pn$(0$^+$), and
$pp$(0$^+$)+$nn$(0$^+$). Here, for example, $^3$H($\frac{1}{2}^+$) stands for not 
only the ground state of $^3$H but also its excited states. The latter 
are actually unbound, and these configurations together with the 
ground-state wave function are obtained by
diagonalizing the intrinsic Hamiltonian for the $p$+$n$+$n$ system 
in ${\cal L}^2$ basis functions. Similarly $pn$(0$^+$), $pp$(0$^+$), 
and $nn$(0$^+$) stand for the two-nucleon pseudo states with the isospin 
$T=1$. 

The total wave function $\Psi^{JM\pi}$ may be expressed in terms of a combination of 
various components, $\sum_{\rm AB}\Psi_{\rm AB}^{JM\pi}$, with 
\begin{equation}
\Psi_{\rm AB}^{JM\pi} = 
\sum_{i=1}^{N_A}  \sum_{j=1}^{N_B} \sum_{I, \ell}
{\cal A}
[[\Phi^{{\rm A}, i}_{J_A \pi_A}\Phi^{{\rm B}, j}_{J_B \pi_B}]_I \; 
\chi_c]_{JM},
\end{equation}
where e.g. $N_A$ is the 
basis size for the nucleus A, $\Phi^{{\rm A},i}_{J_A \pi_A}$ is the
intrinsic wave function of its $i$th state with the angular momentum
$J_A$ and the parity $\pi_A$, and $\chi_c$ 
is the relative motion function between the two nuclei. 
The angular momenta of the two nuclei are coupled to the channel 
spin $I$, which is further coupled with 
the partial wave $\ell$ for the relative motion to 
the total angular momentum $JM$. The index $c$ 
denotes a set of ($i$, $j$, $I$, $\ell$). The parity $\pi$ of the 
total wave function is $\pi_A \pi_B (-1)^{\ell}$. 

In the MRM the configuration space is divided into two regions,
internal and external, by a channel radius. The total wave function 
in the internal region, $\Psi_{\rm int}^{JM\pi}$, is constructed by 
expanding $\chi_c(\vi r)$ in terms of 
$r^{\ell}{\rm exp}(-\rho r^2)Y_{\ell}(\hat{\vi r})$
with a suitable set of $\rho$, 
while the total wave function in the external region, 
$\Psi_{\rm int}^{JM\pi}$, is represented by expressing $\chi_c$ with 
Coulomb or Whitakker functions 
depending on whether the channel is open or not. 
The scattering wave function and the $S$-matrix 
are determined by solving a Schr\"odinger equation 
\begin{equation}
[ H - E + \widetilde{L}] \Psi_{\rm int}^{JM\pi} = \widetilde{L} \Psi_{\rm ext}^{JM\pi} 
\end{equation}
in the internal region together with the continuity 
condition $\Psi_{\rm int}^{JM\pi}=\Psi_{\rm ext}^{JM\pi}$ at
the channel radius. Here $\widetilde{L}$ is the Bloch operator. See
Ref.~\cite{desc} for detail. 

In the MRM the ground-state wave function of
$^4$He is approximated in 
combinations of the multi-channel configurations.

\section{Model}

\label{model.sec}

\subsection{Hamiltonian}

The Hamiltonian we use reads 
\begin{align}
H=\sum_{i} T_i-T_\text{cm}+\sum_{i<j}v_{ij}+\sum_{i<j<k}v_{ijk}.
\end{align}
The  kinetic energy of the center of mass motion 
is subtracted and the two-nucleon interaction $v_{ij}$ 
consists of nuclear and Coulomb parts. 
As the $NN$ potential we employ 
Argonne $v$8$^\prime$ (AV8$^{\prime}$)~\cite{AV8p} 
and G3RS~\cite{tamagaki} potentials that  
contain central, tensor and spin-orbit components.
The $\bm{L}^2$ and $(\bm{L}\cdot\bm{S})^2$ terms in the G3RS potential
are omitted. The $NN$ potential of AV8$^{\prime}$ type 
contains eight pieces: $v_{ij}$=$\sum_{p=1}^8v^{(p)}(r_{ij})
{\cal O}_{ij}^{(p)}$, where $v^{(p)}(r_{ij})$ and  ${\cal O}_{ij}^{(p)}$
are the radial form factor and the operator characterizing 
each piece of the potential. The operators are defined as 
${\cal O}_{ij}^{(1)}$=1,\ ${\cal O}_{ij}^{(2)}$=${\vi \sigma}_i\cdot{\vi
\sigma}_j$,\ ${\cal O}_{ij}^{(3)}$=${\vi \tau}_i\cdot{\vi
\tau}_j$,\ ${\cal O}_{ij}^{(4)}$=${\vi \sigma}_i\cdot{\vi
\sigma}_j {\vi \tau}_i\cdot{\vi \tau}_j$,\ ${\cal O}_{ij}^{(5)}$=$S_{ij}$,
\ ${\cal O}_{ij}^{(6)}$=$S_{ij}{\vi \tau}_i\cdot{\vi \tau}_j$,\ 
${\cal O}_{ij}^{(7)}$=$({\vi L}\cdot{\vi S})_{ij}$,\ ${\cal
O}_{ij}^{(8)}$=$({\vi L}\cdot{\vi S})_{ij}{\vi \tau}_i\cdot{\vi
\tau}_j$, where $S_{ij}$ is the tensor operator, and 
$({\vi L}\cdot{\vi S})_{ij}$ is the spin-orbit operator. 
For the sake of later convenience, we define $V_p$ by
\begin{align}
V_p=\sum_{i<j}v^{(p)}(r_{ij}){\cal O}^{(p)}_{ij}.
\end{align}

The AV8$^\prime$ potential is more repulsive 
at short distances and has a stronger tensor component than the G3RS potential.
Due to this property one has to perform calculations of 
high accuracy 
particularly when the AV8$^\prime$ potential is used, in order to 
be safe from those problems of the CSM that are raised by Wita{\rm \l}a and 
Gl\"{o}ckle~\cite{witala}.
To reproduce the two- and three-body threshold energies is vital for 
a realistic calculation of $\sigma_{\gamma}(E_{\gamma})$. 
To this end we add 
a three-nucleon force (3NF) $v_{ijk}$, and adopt a purely phenomenological 
potential~\cite{hiyama} that is determined to fit the 
inelastic electron form factor
from the ground state to the first excited state of $^{4}$He as well as 
the binding energies of $^3$H, $^3$He and $^4$He. 

\subsection{Gaussian basis functions}
\label{basisfunction}

Basis functions defined here can apply to any number $N$ of nucleons. 
The basis function we use for $N$-nucleon system takes a general form in $LS$ coupling scheme
\begin{align}
\Phi_{(LS)JMTM_T}^{(N)\pi}=\mathcal{A}\left[\phi_L^{(N)\pi}
\chi_S^{(N)}\right]_{JM}
\eta_{TM_T}^{(N)},
\label{LScoupling}
\end{align}
where $\mathcal{A}$ is the antisymmetrizer. 
We define spin functions by a successive coupling of each spin function $\chi_{\frac{1}{2}}(i)$ 
\begin{align}
&\chi_{S_{12}S_{123}\dots S M_S}^{(N)}\notag\\
&= [\dots[[\chi_{\frac{1}{2}}(1)\chi_{\frac{1}{2}}(2)]_{S_{12}}
\chi_{\frac{1}{2}}(3)]_{S_{123}}\dots]_{SM_S}.
\label{spin.func}
\end{align}
Since taking all
possible intermediate spins ($S_{12}, S_{123}$, $\dots$) forms a
complete set for a given $S$, any spin function $\chi_S^{(N)}$ 
can be expanded in terms of the functions~(\ref{spin.func}). 
Similarly the isospin 
function $\eta_{TM_T}^{(N)}$ can also be expanded using 
a set of isospin functions $\eta_{T_{12}T_{123}\dots TM_T}^{(N)}$. 
In the MRM calculation we use 
a particle basis that in general contains a mixing of the 
total isospin $T$, which is caused by the Coulomb potential. 
 
There is no complete set that is flexible enough to describe 
the spatial part $\phi_L^{(N)\pi}$. For example, 
harmonic-oscillator functions are quite inconvenient to describe spatially 
extended configurations. We use an expansion in terms of correlated 
Gaussians~\cite{varga,svm}. 
As demonstrated in Ref.~\cite{kamada}, the Gaussian basis 
leads to accurate solutions for few-body bound states interacting 
with the realistic $NN$ potentials. 

Two types of Gaussians are used. One is a basis expressed in 
a partial wave expansion
\begin{align}
&\phi_{\ell_1 \ell_2(L_{12})\ell_3(L_{123})\dots LM_L}^{(N)\pi}(a_1, a_2,
 \dots, a_{N-1})
 \notag\\
&=\exp(-a_1{x}_1^2-a_2{x}_2^2-\dots-a_{N-1}{x}_{N-1}^2)\notag\\
& \times 
 [\dots[[{\cal Y}_{\ell_1}(\vi x_1){\cal Y}_{\ell_2}(\vi x_2)]_{L_{12}}
{\cal Y}_{\ell_3}(\vi x_3)]_{L_{123}}\dots]_{LM_L}
\label{PWE}
\end{align}
with
\begin{equation}
{\cal Y}_{\ell}(\vi r)=r^{\ell}Y_{\ell}(\hat{\vi r}).
\end{equation}
Here the coordinates ${\vi x}_1,\, {\vi x}_2,\, \dots, {\vi x}_{N-1}$
are a set of relative coordinates. The angular part is represented by 
successively coupling the partial wave $\ell_i$ associated with each coordinate. 
The values of $a_i$ and $\ell_i$ as well as the intermediate angular
momenta $L_{12},\, L_{123}, \dots$ are variational parameters. 
The angular momentum $\ell_i$ is limited to 
$\ell_i \leq 2$ in the present calculation. 
This  basis is employed to construct the 
internal wave function $\Psi^{JM\pi}_{\rm int}$ 
of the MRM calculation. 

The other is an explicitly correlated Gaussian with 
a global vector representation~\cite{varga,svm,dgvr,fbaoyama}
\begin{align}
&\phi_{L_1 L_2(L_{12})L_3 LM_L}^{(N)\pi}(A, u_1,
 u_2, u_3)\notag\\
&=\exp(-\tilde{\bm{x}}A\bm{x})
 [[\mathcal{Y}_{L_1}(\tilde{u}_1\bm{x})
\mathcal{Y}_{L_2}(\tilde{u}_2\bm{x})]_{L_{12}}
{\cal Y}_{L_3}(\tilde{u}_3\bm{x})]_{LM_L},
\label{GVR.eq}
\end{align}
where $A$ is an $(N\!-\!1)\times (N\!-\!1)$ positive definite symmetric matrix
and $u_i$ is an $(N-1)$-dimensional column vector. Both $A$ and
$u_i$ are variational parameters. 
The tilde symbol denotes a transpose, that is,  
$\tilde{\bm{x}}A\bm{x}=\sum_{i,j=1}^{N-1}A_{ij}\bm{x}_i\cdot\bm{x}_j$ 
and $\tilde{u}_i\bm{x}=\sum_{k=1}^{N-1}(u_i)_k\bm{x}_k$. The latter
specifies the global vector $u_i$ responsible for the rotation.
The basis function~(\ref{GVR.eq}) will be used in the CSM calculation.
Actually a choice of the angular part of 
Eq.~(\ref{GVR.eq}) is here restricted to $L_3=0$. With the two global 
vectors any $L^{\pi}$ states but $0^-$ 
can be constructed with a suitable 
choice of $L_1$ and $L_2$. 

Apparently the basis function~(\ref{GVR.eq}) includes correlations among 
the nucleons through the non-vanishing off diagonal elements of $A$. Contrary
to this, the basis function~(\ref{PWE}) takes a product form of a function
depending on each
coordinate, so that the correlation is usually accounted for by including 
the so-called rearrangement channels that are described 
with different coordinate 
sets~\cite{kamimura}. A great advantage of Eq.~(\ref{GVR.eq}) is that 
it keeps its functional form under the coordinate transformation. Hence  
one needs no such rearrangement channels but can use just one 
particular coordinate set, which 
enables us to calculate Hamiltonian matrix elements 
in a unified way. See Refs.~\cite{dgvr,fbaoyama} 
for details. 

The variational parameters are determined 
by the stochastic variational method~\cite{varga,svm}.  It is confirmed 
that both types of basis functions produce accurate results for 
the ground-state properties of $^3$H, $^3$He, and $^4$He~\cite{dgvr}.   
Table~\ref{spectrum3bf} lists the properties of $^3$H and $^4$He
obtained using the basis~(\ref{GVR.eq}). 
Included $L_1$ and $L_2$ values are the same 
as those used in Refs.~\cite{dgvr,inversion}. Both potentials of 
AV8$^{\prime}$+3NF and G3RS+3NF reproduce the binding energy and the root-mean square 
radius of $^4$He satisfactorily. The G3RS+3NF potential gives a 
slightly larger radius and a smaller $D$-state probability $P(2, 2)$ 
than the AV8$^{\prime}$+3NF potential. 

\begin{table}
\caption{Ground-state properties of $^3$H and $^4$He 
calculated with the correlated Gaussians~(\ref{GVR.eq})  
using the AV8$^\prime$ and G3RS potentials together with 3NF. 
Here $E$, $\sqrt{\left<r_p^2\right>}$, and $\sqrt{\left<r_{pp}^2\right>}$ 
denote the energy, the root-mean-square radius of proton distribution and 
the root-mean-square relative distance of protons, respectively, 
and $P(L, S)$ stands for the probability (in \%) of finding the component with 
the total orbital angular momentum $L$ and the spin $S$. 
The experimental energy of $^4$He is $-28.296$\,MeV and the 
point proton radius is 1.457(14)\,fm~\cite{mueller}. }
\label{spectrum3bf}
\begin{center}
\begin{tabular}{ccccccccc}
\hline\hline
         &&\multicolumn{3}{c}{AV8$^\prime$+3NF}&&\multicolumn{3}{c}{G3RS+3NF}\\
\cline{3-5}\cline{7-9}
         &&$^3$H &&$^4$He&&$^3$H &&$^4$He\\
\hline
$E$\ (MeV)      &&$-$8.41&&$-$28.43&&$-$8.35 &&$-$28.56\\
$\sqrt{\left<r_p^2\right>}$\ (fm) &&1.70&&1.45&&1.74&&1.47\\
$\sqrt{\left<r_{pp}^2\right>}$\ (fm) && -- && 2.41 && -- && 2.45 \\
$P(0, 0)$&&91.25&&85.56&&92.85&&88.33\\
$P(2, 2)$&& 8.68&&14.07&& 7.10&&11.42\\
$P(1, 1)$&& 0.07 && 0.37 && 0.05&& 0.25\\
\hline\hline
\end{tabular}
\end{center}
\end{table}

\subsection{Two- and three-body decay channels}
\label{decay}

As is well-known, the electric dipole operator 
\begin{align}
\mathcal{M}_{1\mu}&=
\sum_{i=1}^4\frac{e}{2}(1-\tau_{3_i})(\bm{r}_i-\bm{x}_4)_\mu \nonumber \\
&=-\frac{e}{2}\sqrt\frac{4\pi}{3}
\sum_{i=1}^4 \tau_{3i}\mathcal{Y}_{1\mu}(\bm{r}_i-\bm{x}_4) \nonumber \\
&=\frac{e}{2}\sqrt\frac{4\pi}{3}\Big\{
\frac{1}{2}(\tau_{3_1}-\tau_{3_2})\mathcal{Y}_{1\mu}(\bm{x}_1)\nonumber \\
&\qquad +\frac{1}{3}(\tau_{3_1}+\tau_{3_2}-2\tau_{3_3})\mathcal{Y}_{1\mu}(\bm{x}_2)\nonumber \\
&\qquad +\frac{1}{4}(\tau_{3_1}+\tau_{3_2}+\tau_{3_3}-3\tau_{3_4})\mathcal{Y}_{1\mu}(\bm{x}_3)\Big\}
\label{dipole.eq}
\end{align}
is an isovector, where $\bm{x}_4$ is the center of mass coordinate 
of $^4$He, and ${\bm{x}_i}$ is the Jacobi coordinate: $\bm{x}_1=\bm{r}_2-\bm{r}_1$, 
$\bm{x}_2=\bm{r}_3-\frac{1}{2}(\bm{r}_1+\bm{r}_2)$, $\bm{x}_3=\bm{r}_4-
\frac{1}{3}(\bm{r}_1+\bm{r}_2+\bm{r}_3)$. This operator excites the ground state of $^4$He to 
those states that have $J^{\pi}T$=$1^-1$ in so far as 
a small isospin admixture in the ground state of $^4$He is ignored. 
Moreover those excited states 
should mainly have $(L,S)=(1,0)$ component, because the ground 
state of $^4$He is dominated by the 
$(0,0)$ component. See Table~\ref{spectrum3bf}. 
Excited states with $S=1$ or 2 components will be 
weakly populated by the $E1$ transition through the minor 
components (12--14\%) of the $^4$He ground state.

According to the $R$-matrix phenomenology as quoted in Ref.~\cite{tilley}, 
two levels with $1^-1$ are identified. Their excitation energies and
widths in MeV are respectively $(E_{\rm x},\, \Gamma)$=(23.64,\,
6.20), (25.95,\, 12.66). We have recently 
studied the level structure of $^4$He and succeeded to reproduce all 
the known levels below 26 MeV~\cite{inversion}. With including 
the 3NF, two $1^-1$ states are predicted at about 23 and 27 MeV in case 
of the AV$8^{\prime}$ potential. They are however 
not clearly identified as resonances in a recent microscopic scattering
calculation~\cite{fbaoyama}. In Sec.~\ref{eds.transition}, we will 
show that three states with strong $E1$ strength are obtained 
below 35\,MeV in a diagonalization using the ${\cal L}^2$ basis 
and will discuss the properties of those states.  

Low-lying excited states with $1^-1$ decay to $^3$H+$p$ and $^3$He+$n$
channels with $P$ wave. Possible channel spins
$^{2I+1}\ell_{J}$ that the $^3$H+$p$ or $^3$He+$n$ continuum state takes  
are $^1P_1$ and $^3P_1$~\cite{fbaoyama}. 
A main component of the $^1P_1$ continuum state is found to be 
$(L,S)=(1,0)$ while that of the $^3P_1$ continuum state is 
$(1,1)$. Thus it is expected that the $E1$ excitation of $^4$He 
is followed mainly by
the $^3$H+$p$ and $^3$He+$n$ decays in the $^1P_1$
channel, which agrees with the result of a resonating 
group method calculation including the $^3$H+$p$, $^3$He+$n$, and 
$d$+$d$ physical channels~\cite{wachter}. 

The two-body decay to $d$+$d$ is suppressed due to the isospin
conservation. Above the $d$+$p$+$n$ threshold at
26.07\,MeV, this three-body decay becomes possible where the decaying 
$pn$ pair is in the $T=1$ state. In fact the cross section to 
this three-body decay is observed experimentally~\cite{shima}.

\subsection{Square-integrable basis with $J^{\pi}T=1^-1$}
\label{squareintegrable}

\begin{figure}[ht]
\begin{center}
\epsfig{file=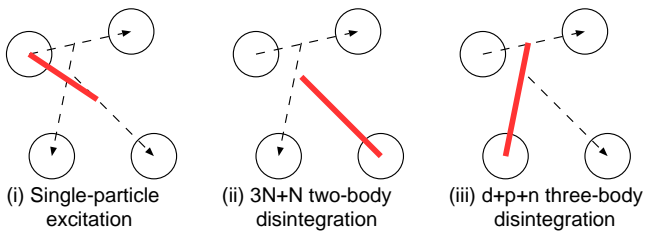,scale=0.76}
\caption{(Color online) Three patterns for the dipole excitations for
 $^4$He. Thick solid lines denote the coordinates on which the spatial 
part of the $E1$ operator acts.}
\label{dipole_excit.fig}
\end{center}
\end{figure}

The accuracy of the CSM calculation crucially depends on how sufficiently 
the ${\cal L}^2$ basis functions $\Phi_i({\bm x})$ for $J^{\pi}T=1^-1$ 
are prepared for solving 
the eigenvalue problem~(\ref{ceig.eq}). We attempt at constructing the 
basis paying attention to two points: the sum rule 
of $E1$ strength and the decay channels as discussed 
in Sec.~\ref{decay}. 
As the $E1$ operator~(\ref{dipole.eq}) suggests, we will construct the 
basis with $1^-$ by 
choosing the following three operators and acting them on the basis 
functions that constitute the ground state of $^4$He:  
(i) a single-particle excitation built with $\mathcal{Y}_{1\mu}(\bm{r}_1-\bm{x}_4)$, 
(ii) a $3N$+$N$ ($^3$H+$p$ and $^3$He+$n$)  
two-body disintegration due to $\mathcal{Y}_{1\mu}(\bm{x}_3)$, 
(iii) a $d$+$p$+$n$ three-body disintegration due to $\mathcal{Y}_{1\mu}(\bm{x}_2)$. 
See Fig.~\ref{dipole_excit.fig}.   
The basis (i) is useful for satisfying the sum rule, 
and the bases (ii) and (iii) take care of the two- and 
three-body decay asymptotics. These cluster configurations 
will be better described using the relevant coordinates 
rather than the single-particle coordinate. It should be noted that 
the classification label does not necessarily indicate
strictly exclusive meanings 
because the basis functions belonging to the different classes 
have some overlap among others because of their non-orthogonality. 

We will slightly truncate the ground-state wave functions of 
$^3$H, $^3$He, and $^4$He when they are needed to construct the above 
$1^-$ configurations, (i) and (ii). With this truncation 
a full calculation 
presented in Sec.~\ref{results.sec} will be possible without 
excessive computer time. As shown in 
Table~\ref{spectrum3bf}, the ground 
states of these nuclei contain a small amount (less than 0.5\%) 
of $L=1$ component, so that we omit this component and 
reconstruct the ground-state wave functions 
using only with $L_1$=0, 2 and $L_2=L_3=0$ in Eq.~(\ref{GVR.eq}). 
The energy loss is found to be small compared to the accurate energy of 
Table~\ref{spectrum3bf}. E.g., in the case of AV8$^{\prime}$+3NF, 
the loss is 0.23 MeV  for $^{3}$H in 64 basis dimension 
and 1.53 MeV for $^{4}$He in 200 basis dimension. 
The truncated ground-state wave function is denoted 
$\Psi_{\frac{1}{2}M\frac{1}{2}M_T}^{(3)}$ for $3N$ and 
$\Psi_{0000}^{(4)}$ for $^4$He. 

Note, however, that we use the accurate wave function $\Psi_0$ 
of Table~\ref{spectrum3bf} for the  $^4$He ground state 
in computing $S(E)$ with Eq.~(\ref{csm.strength}).

\subsubsection{Single-particle (sp) excitation}
\label{s.p.excit}

As is well-known, applying the $E1$ operator on a ground state 
leads to a coherent state that exhausts all the $E1$ strength from the 
ground state. The coherent state is however not an 
eigenstate of the Hamiltonian. 
In analogy to this, the basis of type (i) 
is constructed as follows
\begin{align}
\Psi_{f}^{\text{sp}}
&=\mathcal{A}\left[\Phi_{0}^{(4)}(i)
\mathcal{Y}_{1}(\bm{r}_1-\bm{x}_4)\right]_{1M}\eta_{T_{12}T_{123}10}^{(4)},
\label{EDbasis}
\end{align}
where $\Phi_{0}^{(4)}(i)$ is the space-spin part of the $i$th basis function 
of $\Psi_{0000}^{(4)}$. We include all the basis functions  
and all possible $T_{12},
T_{123}$ for the four-nucleon isospin state with $TM_T$=10. 
The truncated basis $\Phi_{0}^{(4)}(i)$ consists of either
$[\phi^{(4)+}_0\chi_0^{(4)}]_{0}$ or 
$[\phi^{(4)+}_2\chi_2^{(4)}]_{0}$ in the notation of 
Eq.~(\ref{LScoupling}). The former
contains no global vector, while the latter contains one global vector. 
Since $\mathcal{Y}_{1}(\bm{r}_1-\bm{x}_4)$ is rewritten as 
$\mathcal{Y}_{1}(\tilde{w}\bm{x})$ with $\tilde{w}=(-\frac{1}{2}, 
-\frac{1}{3},-\frac{1}{4})$,  
the basis~(\ref{EDbasis}) contains at most two global vectors and 
reduces to the 
correlated Gaussian~(\ref{GVR.eq}). For example, the basis with the
latter case 
can be reduced, after the angular momentum recoupling, to the standard
form with $L_1=2,\, L_2=1$
\begin{align}
&\left[[\phi^{(4)+}_{20(2)02}(A,u_1)\chi_{1\frac{3}{2}2}^{(4)}]_0
\mathcal{Y}_{1}(\bm{r}_1-\bm{x}_4)\right]_{1M}\nonumber \\
&=\sum_{L_{12}=1,2,3}\sqrt{\frac{2L_{12}+1}{15}}
\left[\phi^{(4)-}_{21(L_{12})0L_{12}}(A,u_1,w)\chi_{1\frac{3}{2}2}^{(4)}\right]_{1M}.
\end{align}
Each $L_{12}$ component of
$[\phi^{(4)-}_{21(L_{12})0L_{12}}(A,u_1,w)\chi_{1\frac{3}{2}2}^{(4)}]_{1M}$ is 
included as an independent basis function in what follows.

\subsubsection{$3N$+$N$ two-body disintegration}

In this basis the nucleon couples with the ground and pseudo 
states of the $3N$
system. Their relative motion carries 
$P$-wave excitations, and it is described in a combination of 
several Gaussians. The basis function takes the following form 
\begin{align}
\Psi_{f}^{\text{3NN}}
&=\mathcal{A}\left[\Phi^{(3)}_{J_3}(i)
\exp\left(-a_3{x_3^2}
\right)[\mathcal{Y}_{1}(\bm{x}_3)
\chi_{\frac{1}{2}}(4)]_j\right]_{1M} \notag\\
&\times[\eta_{T_{12}\frac{1}{2}}^{(3)}\eta_{\frac{1}{2}}(4)]_{10},
\end{align}
where $\Phi_{J_3}^{(3)}(i)$ is the  space-spin part of 
the $i$th basis function of $\Psi_{\frac{1}{2}M\frac{1}{2}M_T}^{(3)}$. 
The value of $j$ takes  $\frac{1}{2}$ and $\frac{3}{2}$, and 
$J_3$ takes any of $\frac{1}{2}, \, \frac{3}{2}$, and $\frac{5}{2}$ that, 
with $j$, can add up to the angular momentum 1. The parameter $a_3$ 
is taken in a geometric progression as 
$12.5/1.4^{2(k-1)} \,(k=1,2,\ldots, 15)$ in fm$^{-2}$. 
As in the basis of the single-particle excitation 
the space-spin part is again expressed
in the correlated Gaussians~(\ref{GVR.eq}) with at most two global
vectors, where one of the global vectors is 
$\mathcal{Y}_{1}(\bm{x}_3)$=$\mathcal{Y}_{1}(\tilde{w}\bm{x})$ with 
$\tilde{w}=(0,0,1)$. All the basis states with different values of 
$J_3$ and $j$ are included independently.

\subsubsection{$d$+$p$+$n$ three-body disintegration}

In this basis the relative motion between $3N$ and $N$ is 
$S$ wave but 
the $3N$ system is excited to the $d$+$N$ configuration with 
$P$-wave relative motion. 
Here $d$ does not necessarily mean its ground
state but include pseudo states with the 
angular momentum $J_2^{\pi}=0^+,1^+,2^+,3^+$. The
spatial part is however taken from the basis functions of the deuteron 
ground state. The three-body basis function takes the following form 
\begin{align}
\Psi_{f}^{\text{dpn}}&=
\mathcal{A}\left[\Phi_{J_3}^{(\text{dN})}(i)
\exp\left(-a_3 x_3^2\right)
[\mathcal{Y}_{0}(\bm{x}_3)\chi_{\frac{1}{2}}(4)]_{\frac{1}{2}}\right]_{1M}\notag\\
&\times[\eta_{0\frac{1}{2}}^{(3)}(123)\eta_{\frac{1}{2}}(4)]_{10},
\label{dpn.exc}
\end{align}
with
\begin{align}
\Phi_{J_3}^{(\text{dN})}(i)
=\left[\Psi_{J_2}^{(2)}(i)
\exp\left(-a_2{x_2^2}
\right)[\mathcal{Y}_{1}(\bm{x}_2)
\chi_{\frac{1}{2}}(3)]_{j}\right]_{J_3},
\end{align}
where $\Psi_{J_2}^{(2)}(i)$ is the (pseudo) deuteron wave function
mentioned above. Both of $J_3$ and $j$ take 
$\frac{1}{2}$ and $\frac{3}{2}$. 
All possible sets of $J_3,\, J_2$ and $j$ values that
satisfy the angular momentum addition rule are included in the calculation. 
Both $a_2$ and $a_3$ are again given in a geometric progression, 
$3.125/1.4^{2(k-1)}\, (k=1,2,\ldots, 10)$ in fm$^{-2}$. 
Note that 
$\mathcal{Y}_{1}(\bm{x}_2)$=$\mathcal{Y}_{1}(\tilde{w}\bm{x})$ with 
$\tilde{w}=(0,1,0)$. After recoupling the orbital and spin angular
momenta, the basis~(\ref{dpn.exc}) leads to the following space-spin 
parts: $[\phi^{(4)-}_{L_1 1(L)0L}\chi_{1S_{123}S}^{(4)}]_{1M}$ with 
$L_{1}$=0 or 2, and all possible values of $L, \, S_{123},\, S$ are 
allowed. These are included independently. Note that the matrix $A$ of 
$\phi^{(4)-}_{L_11(L)0L}$ becomes diagonal.

The basis dimension included is 7400 (7760) 
for AV8$^{\prime}$ (G3RS)+3NF,   
1200 (1560) from (i), 3000 from (ii), 
and 3200 from (iii), respectively. 

\section{Results}

\label{results.sec}

\subsection{Discretized  strength of electric dipole transition}
\label{eds.transition}

\begin{figure}[ht]
\begin{center}
\epsfig{file=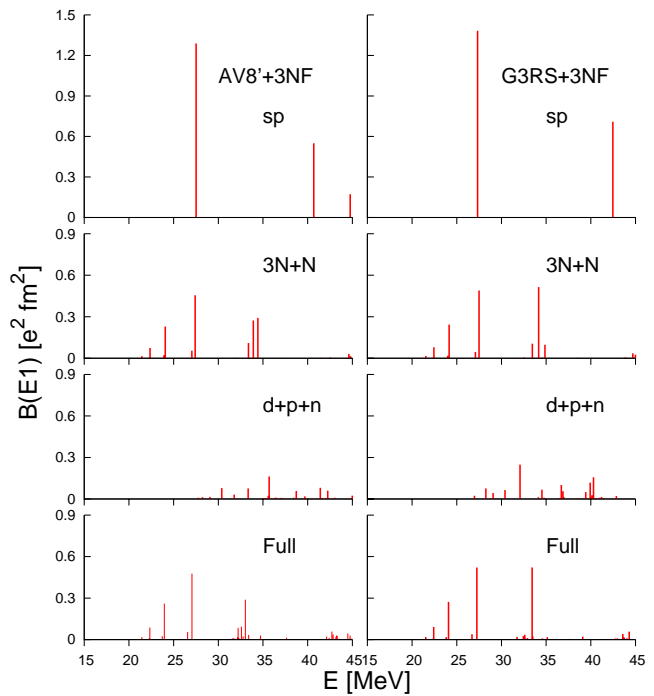,scale=0.78}
\caption{(Color online) Discretized strength of the $E1$ 
transitions in $^4$He.  
See the text for the calculations classified by sp, $3N$+$N$, 
$d$+$p$+$n$, and Full. }
\label{E1G3RS.fig}
\end{center}
\end{figure}

Continuum states with $J^{\pi}T=1^-1$ are discretized by diagonalizing 
the Hamiltonian in the basis functions defined in Sec.~\ref{model.sec}. 
These discretized states provide us with an approximate distribution 
of the $E1$ strength. 
Figure~\ref{E1G3RS.fig} displays the reduced transition probability 
\begin{align}
B(E1,\lambda)=\sum_{M\mu}\left|\left<\Psi_{\lambda}^{1M-}(\theta=0)|\mathcal{M}_{1\mu}
|\Psi_{0}\right>\right|^2.
\label{B(E1)}
\end{align}
as a function of the discretized energy 
$E_{\lambda}(\theta =0)$. The calculations were performed in each 
basis set of (i)--(iii) as well as a full basis that includes all of them. 
The distribution of $B(E1,\lambda)$ depends rather weakly on the potentials.

As expected, three types of basis functions play a distinctive and 
supplementary role in the $E1$ strength distribution. 
The basis functions (i) produce 
strongly concentrated strength at about 27\,MeV and 
another peak above 40 MeV. The $(L,S)=(1,0)$ 
component of these states is about 95\%.  
With the $3N$+$N$ two-body configurations (ii), we obtain two peaks 
in the region of 20--30\,MeV 
and one or two peaks at around 35\,MeV. 
The two peaks at about 25 MeV may perhaps correspond to 
the $1^-$ levels with $T$=1 at 23.64 and 25.95 MeV with very broad 
widths~\cite{tilley}. Note, however, that 
a microscopic four-nucleon scattering 
calculation presents no conspicuous resonant phase shifts 
for $^3P_1$ and $^1P_1$ channels~\cite{fbaoyama}. 
The three-body configurations (iii) give relatively small strength 
broadly in the excitation energy above 30 MeV. 
The three prominent peaks at around 25--35\,MeV remain to exist 
in the full basis calculation. 
This implies that the low-lying strength 
mainly comes from the $3N$+$N$ configuration. 
We will return this issue in Sec.~\ref{comparison}
The three discretized states are labeled by their excitation
energies $E_i$ in what follows. 

Table~\ref{E1.peak} shows the properties of the three states 
$E_i$ that have strong strength. 
The expectation value of each piece of 
the Hamiltonian is a measure of its contribution to the energy. 
We see that the central  
($V_4$: ${\vi \sigma}_i\cdot{\vi \sigma}_j{\vi \tau}_i\cdot{\vi \tau}_j$) 
and tensor ($V_6$: $S_{ij}{\vi \tau}_i\cdot{\vi \tau}_j$) terms 
are major contributors among the interaction pieces. 
The one-pion exchange potential (OPEP) consists of $V_6$ and $V_4$
terms, so that the tensor force of the OPEP is found to play a vital role. 
The value of $P(L, S)$ in the table is 
obtained by the squared coefficient $(C_{LS}^\lambda)^2$ of the expansion 
\begin{align}
\Psi_\lambda^{1M-}(\theta=0)=\sum_{LS}C_{LS}^\lambda\Psi_{(LS)1M10}^{(4)-},
\end{align}
where $\Psi_{(LS)1M10}^{(4)-}$ is normalized. 
Note that no basis functions with $L^{\pi}=0^-$ are 
included in the present calculation as they are not expressible in the
two global vectors. As expected, all of the three states 
dominantly consist of the $(1, 0)$ component, which can be excited, 
by the $E1$ operator, from the main 
component $(0, 0)$ of the $^4$He ground state.
We see a considerable admixture of the $S=2$ components 
especially with  $L=3$ in the three states.
This is understood from the role played by the tensor force that 
couples the $S=0$ and 2 states. In fact the $S=2$ states lose energy 
due to large kinetic energy contributions but gain energy owing to the 
coupling with the main component with $(1,0)$ through the tensor force. 
For example, in the case of $E_1$ state, the diagonal matrix elements 
of the kinetic energy,
$\langle\Psi_{(L2)1M10}^{(4)-}|T|\Psi_{(L2)1M10}^{(4)-}\rangle$,
are 196.5 (160.3), 198.6 (161.5), 199.3 (162.3) MeV for $L$=1, 2, 3 
with AV8$^\prime$ (G3RS)+3NF, 
while the tensor coupling 
matrix elements between $(1, 0)$  and $(L, 2)$ states, 
$\langle\Psi_{(10)1M10}^{(4)-}|V_5+V_6|\Psi_{(L2)1M10}^{(4)-}\rangle$, are
respectively $-$54.5 ($-$40.2), $-$70.8 ($-$52.1), $-$84.0 ($-$61.9)  
MeV for $L$=1, 2, 3 states.

\begin{table}
\caption{Properties of the three $1^-1$ states that exhibit strong 
$B(E1)$ strength. The excitation energy $E$ and the 
expectation values are given in units of 
MeV. The value of $P(L, S)$ is given in \%. See Table~\ref{spectrum3bf} for 
the ground-state energy of $^4$He. }  
\label{E1.peak}
\begin{tabular}{lccccccccc}
\hline\hline
             &&\multicolumn{3}{c}{AV8$^\prime$+3NF}&&\multicolumn{3}{c}{G3RS+3NF}\\
\cline{3-5}\cline{7-9}
$E$          &&23.96&27.05&33.02&&24.08&27.25&33.43\\ 
\hline
$\langle H\rangle$&&$-$4.46&$-$1.38&4.60&&$-$4.48&$-$1.31&4.88\\
$\langle T\rangle$&&51.21&54.78&43.71&&44.34&48.37&49.65\\
$\langle V_1\rangle$&&6.42&6.37&4.44&&$-$0.14&$-$0.24&$-$0.31\\
$\langle V_2\rangle$&&$-$3.41&$-$3.68&$-$1.61&&$-$3.07&$-$3.38&$-$2.94\\
$\langle V_3\rangle$&&$-$2.17&$-$2.15&$-$1.65&&$-$3.81&$-$3.75&$-$3.43\\
$\langle V_4\rangle$&&$-$23.83&$-$24.04&$-$16.09&&$-$20.45&$-$20.81&$-$18.46\\
$\langle V_5\rangle$&&0.22&0.22&0.14&&$-$0.41&$-$0.41&$-$0.37\\
$\langle V_6\rangle$&&$-$30.60&$-$30.51&$-$22.71&&$-$20.60&$-$20.64&$-$18.80\\
$\langle V_7\rangle$&&4.79&4.77&3.55&&2.33&2.33&2.13\\
$\langle V_8\rangle$&&$-$6.76&$-$6.73&$-$4.96&&$-$2.37&$-$2.38&$-$2.15\\
$\langle V_{\rm 3NF}\rangle$&&$-$0.74&$-$0.86&$-$0.55&&$-$0.72&$-$0.85&$-$0.85\\
$\langle V_{\rm Coul}\rangle$&&0.42&0.45&0.32&&0.41&0.45&0.42\\
\hline
$P(1, 0)$&&87.18&84.58&82.70&&90.12&88.47&79.73\\
$P(1, 1)$&&4.76 &7.47 &7.59 &&3.18 &4.89&13.86\\
$P(2, 1)$&&0.16 &0.25 &0.22 &&0.09 &0.13&0.36\\
$P(1, 2)$&&0.89 &0.74 &4.56 &&0.85  &0.76&0.95\\
$P(2, 2)$&&2.17 &1.99 &1.40 &&1.89  &1.79&1.41\\
$P(3, 2)$&&4.85 &4.97 &3.53 &&3.86  &3.95&3.69\\
\hline\hline
\end{tabular}
\end{table}

\begin{figure}[ht]
\begin{center}
\epsfig{file=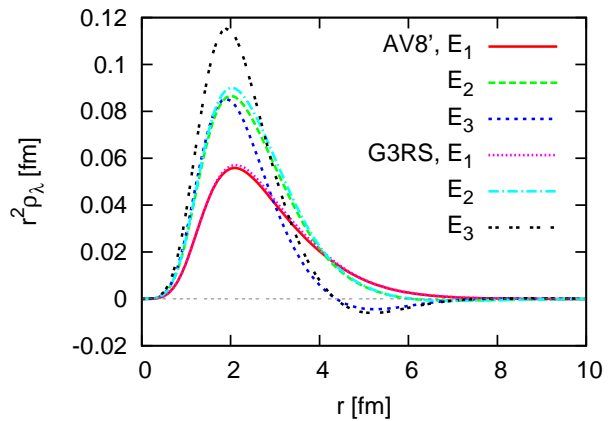,scale=1.3}
\caption{(Color online) Transition densities for the three discretized states 
listed in Table~\ref{E1.peak} that have strong $E1$ strength.}
\label{tran.density.fig}
\end{center}
\end{figure}

The $E1$ transition density is defined as 
\begin{align}
\rho_{\lambda}(r)&=\left<\Psi_{\lambda}^{10-}(\theta =0)\right|\sum_{i=1}^4
\frac{\delta(|\bm{r}_i-\bm{x}_4|-r)}{r^2}\notag\\
&\times\mathcal{Y}_{10}(\bm{r}_i-\bm{x}_4)
\frac{1-\tau_{3_i}}{2}\left|\Psi_0\right>,
\end{align}
which gives the $E1$ transition matrix element through 
\begin{align}
\left<\Psi_{\lambda}^{10-}(\theta
 =0)\right|\mathcal{M}_{10}\left|\Psi_0\right>
=\sqrt{\frac{4\pi}{3}}e\int_{0}^{\infty}\rho_{\lambda}(r) r^2dr.
\end{align}
Figure~\ref{tran.density.fig} displays the transition densities for the 
three states $E_i$ 
of Table~\ref{E1.peak} that give the large $E1$ matrix elements. 
The dependence of
the transition density on the interaction is rather weak
except for the third state labeled by $E_3$. 
The transition density extends
to significantly large distances mainly due to the effect of the $3N$+$N$ 
configurations, so that for a reliable evaluation 
of $B(E1,\lambda)$ the basis functions for $1^-$ must include
configurations that reach far distances. 
The peak of $r^2\rho_{\lambda}(r)$ appears at about 
2\,fm, which is much larger than the peak position (1.1\,fm)  
of $r^2\rho_{\rm g.s.}(r)$, where $\rho_{\rm g.s.}(r)$ is 
the ground-state density of $^4$He. A comparison of the transition 
densities of the second ($E_2$) and third ($E_3$) 
states suggests that near $r\!\approx$\,2-6\,fm  
they exhibit a constructive pattern in the 
second state and a destructive pattern in the third state. 

\subsection{Test of CSM calculation}
\label{test.CSM-MRM}

\begin{figure}[th]
\epsfig{file=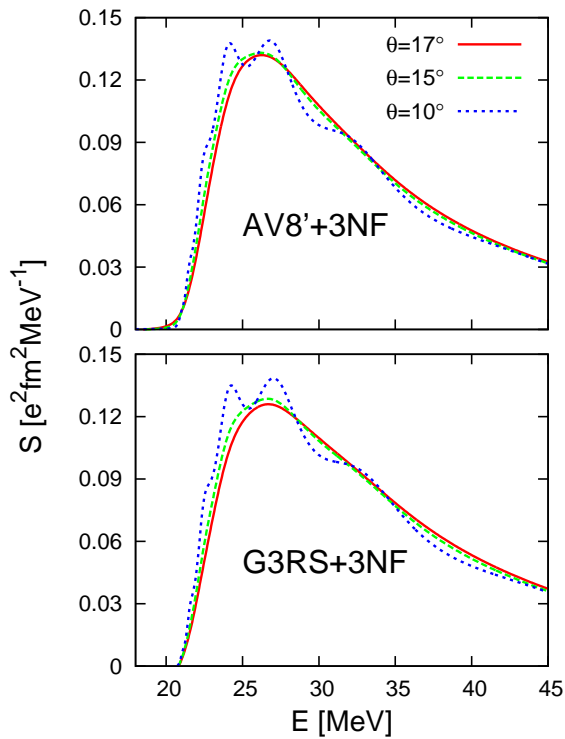,scale=1.2}
\caption{(Color online) Electric dipole strength functions
obtained by the CSM with different rotational angles $\theta$.}
\label{E1CSMtheta.fig}
\end{figure}

The strength function~(\ref{csm.strength}) calculated in the CSM 
using the full basis 
is plotted in Fig.~\ref{E1CSMtheta.fig} for some angles $\theta$. 
Both AV8$^{\prime}$+3NF and G3RS+3NF 
potentials give similar results. 
With $\theta$=10$^\circ$, $S(E)$ shows 
some oscillations whose peaks appear at the energies 
of the discretized states shown in the full calculation of 
Fig.~\ref{E1G3RS.fig}.
To understand this behavior we note that the contribution of an 
eigenstate $\lambda$ to $S(E)$ 
is given by a Lorentz distribution 
\begin{align}
&\frac{1}{\pi}\frac{1}{(E-E_c)^2+\frac{1}{4}\Gamma_c^2}
\sum_{\mu}\Bigg[\frac{1}{2}\Gamma_c
\text{Re}\tilde{\mathcal{D}}^\lambda_\mu(\theta)\mathcal{D}_\mu^\lambda(\theta)
\nonumber \\
&\qquad -(E-E_c)\text{Im}\tilde{\mathcal{D}}^\lambda_\mu(\theta)\mathcal{D}_\mu^\lambda(\theta)
\Bigg],
\end{align} 
where $E_{\lambda}(\theta)=E_c-\frac{i}{2}\Gamma_c$. 
For small angles, $E_c$ is not very different from the 
discretized energy $E_{\lambda}(\theta=0)$ and 
$\Gamma_c$ is small, and therefore  
the strength at $E\approx E_c$ comes mostly from the eigenstate 
$\lambda$ alone because the contribution from the 
neighboring states can be neglected. 
The oscillatory behavior diminishes with increasing $\theta$ or
$\Gamma_c$,   
and finally we obtain one broad peak at 26-27 MeV.
As shown in the figure, the convergence reaches at about 
$\theta$=17$^{\circ}$. 
One might consider $\theta$=17$^{\circ}$ a little too 
small to cover the two $1^-$ states noted in Sec.~\ref{decay}.  
Attempting at including them by increasing $\theta$ 
will lead to numerically unstable and unphysical 
results particularly near the 
$^3$H+$p$ threshold with the present basis dimension.   
Though the strength should in principle vanish 
below the threshold energy, those eigenstates 
$\lambda$ which have large values of $\Gamma_c$ may 
contribute to the strength near the threshold and therefore  
it would be in general hard to obtain vanishing strength 
just below the threshold. 
After some trial and error calculations 
we choose $\theta$=17$^{\circ}$ as an acceptable angle hereafter.

\begin{figure}[ht]
\begin{center}
\epsfig{file=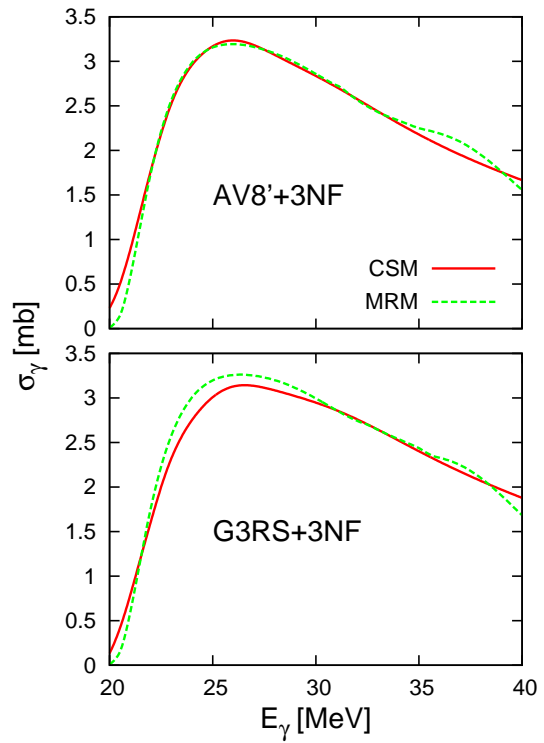,scale=1.2}
\caption{(Color online) Comparison of the 
photoabsorption cross sections calculated with the CSM and the MRM.}
\label{CSM.MRM.fig}
\end{center}
\end{figure}

Figure~\ref{CSM.MRM.fig} compares the photoabsorption cross
sections $\sigma_{\gamma}(E_{\gamma})$ between the CSM and the MRM. 
The calculated $^3$H+$p$ threshold 
is adjusted to agree with experiment. 
As already mentioned, the 
$\sigma_{\gamma}(E_{\gamma})$ value of 
the MRM is defined as a sum of 
$\sigma_{\gamma}^{^3{\rm H}\,p}(E_{\gamma})$ and
$\sigma_{\gamma}^{^3{\rm He}\,n}(E_{\gamma})$. 
Both methods give almost the same cross section,  
which convinces us of the validity of the CSM calculation. 
A little difference appears especially at the energy close to the 
threshold. We think that the reason for that 
is partly because the model space 
employed is not exactly the same each other, partly because 
the MRM calculation does not take into account the three- and four-body 
decay channels and partly because the CSM cross section 
may not be very accurate near the threshold energy as mentioned above. 

A comparison of $\sigma_{\gamma}(E_{\gamma})$ 
between theory and experiment will be made in 
Sec.~\ref{comparison}.

\subsection{Photonuclear sum rules}
\label{E1.sum.rule}

Photonuclear sum rules are related to the moments of 
different order of $\sigma_{\gamma}(E_{\gamma})$. 
The moment is defined as
\begin{align}
m_p(E_{\rm max})=\int_0^{E_{\rm max}} E_{\gamma}^{\ p} 
\sigma_{\gamma}(E_{\gamma})dE_{\gamma}.
\end{align}
The moments 
$m_0,\, m_{-1}$, and  $m_{-2}$ are called the Thomas-Reiche-Kuhn,
bremstrahlungs, and polarizability sum rules, respectively. These  
moments for $E_{\rm max}\to \infty$ 
are expressed with the ground-state expectation values of 
appropriate operators, 
and thus carry interesting electromagnetic 
properties of nuclei~\cite{bohr,lipparini}. 
As is well-known, they are expressed as 
\begin{align}
&m_{-1}(\infty) ={\cal G}\left(Z^2\langle r_p^2\rangle -\frac{Z(Z-1)}{2}\langle r_{pp}^2 \rangle \right),\nonumber \\
&m_0(\infty)={\cal G}\frac{3NZ\hbar^2}{2Am_N}(1+K),
\end{align}
where ${\cal G}=4\pi^2 e^2/3\hbar c$ and $m_N$ is the nucleon mass. 
Here $\langle r_p^2\rangle$ stands for  
the mean square radius of proton distribution 
and $\langle r_{pp}^2 \rangle$ for the
mean square relative distance of protons.  See Table~\ref{spectrum3bf}. 

The enhancement factor $K$ is given as a sum of the 
contributions from the potential pieces, $K$=$\sum_{p=1}^8K_p$, where 
\begin{align}
K_p=\frac{2Am_N}{3NZ\hbar^2 e^2}\frac{1}{2}\sum_{\mu}\langle
\Psi_0|[{\cal M}_{1\mu}^{\dagger},[V_p, {\cal M}_{1\mu}]]|\Psi_0\rangle. 
\end{align} 
For the $NN$ potential of AV8$^{\prime}$ type, $K$ gains contributions from 
the potential piece with the ${\vi \tau}_i\cdot{\vi \tau}_j$
dependence, i.e., the charge-exchange interaction. 
The values of $K_p$ are listed in Table~\ref{Kp.Vp}. For the
sake of reference, we also show the expectation value 
$\langle V_p\rangle$ for the ground state of $^4$He. 
Roughly half of the respective total values, 
$K$ and $\sum_p\langle V_p \rangle$, come  from the tensor 
($V_6$) and central ($V_4$) terms. Since the OPEP contains both $V_6$
and $V_4$ terms, it is instructive to know what values the OPEP predicts 
for $\left<V_p\right>$ and $K_p$. The radial form factor 
$v^{(p)}(r)$ of 
the OPEP is made to vanish for $r\leq 1$\,fm in order to estimate the role 
of the OPEP in the medium- and long-range parts of the $NN$ interaction. 
As shown in Table~\ref{Kp.Vp}, the OPEP 
explains most of the contributions from the $V_6$ term.  
However, the OPEP is not  
enough to account for the $V_4$ contribution. Other central 
forces of $V_4$ type contribute in the medium-range part of the $NN$ interaction. 

The two large contributions originate from the matrix elements for the 
$(L, S)$=(0,0)--(2,2) couplings and the (0,0)--(0,0) diagonal 
channels, respectively. The 
AV8$^{\prime}$ potential has a stronger tensor component than the G3RS
potential, predicting a slightly larger value for $K$. The present value 
of $K$ is smaller than other calculations, e.g.,  
1.14 with the Reid soft core potential~\cite{gari}, 1.29 with the 
AV14+UVII potential~\cite{schiavilla}, and 1.44 with the 
AV18+UIX potential~\cite{gazitb}.

\begin{table}
\caption{Contributions of the eight pieces $V_p$ of the $NN$ potential to the 
enhancement factor $K$ and to the ground state energy of $^4$He given 
in MeV. The values in parentheses are contributions of 
the OPEP that are calculated as explained in the text. 
The $D$-state probability of the deuteron
 is 5.77\% for AV8$^{\prime}$ and 4.78\% for G3RS.} 
\label{Kp.Vp}
\begin{tabular}{ccccccccc}
\hline\hline
         &&\multicolumn{3}{c}{AV8$^\prime$+3NF}&&\multicolumn{3}{c}{G3RS+3NF}\\
\cline{3-5}\cline{7-9}
$p$        &&$\left<V_p\right>$&&$K_p$&&$\left<V_p\right>$&&$K_p$\\
\hline
1&&17.39&& 0 &&1.07 &&0\\
2&&$-$9.59&& 0 &&$-$8.75&&0\\
3&&$-$5.22&&0.011&&$-$9.11&&0.059\\
4&&$-$59.42&&0.460&&$-$51.80&&0.474\\
&&($-$12.51)&&(0.187)&&($-$12.50) &&(0.191) \\
5&&0.75&&0&&$-$0.93&&0\\
6&&$-$70.93&&0.574&&$-$47.16&&0.484\\
 &&($-$68.65)&&(0.667)&& ($-$59.37)&&(0.610) \\
7&&11.09&& 0&& 5.53&& 0\\
8&&$-$15.93&& 0.061 && $-$5.65&&0.025 \\
\hline
Total&& $-$131.9 && 1.11&& $-$116.8&& 1.04\\
\hline\hline
\end{tabular}
\end{table}

The continuum discretized $1^-1$ states calculated with
$\theta=0^{\circ}$ satisfy the sum rule for $m_{-1}(\infty)$ 
almost perfectly: 99.6\% for AV8$^{\prime}$+3NF and 99.7\% 
for G3RS+3NF. This implies that the present basis functions 
sufficiently span the configuration space needed 
to account for all the strengths of the $E1$ transition. 

\begin{figure}[ht]
\begin{center}
\epsfig{file=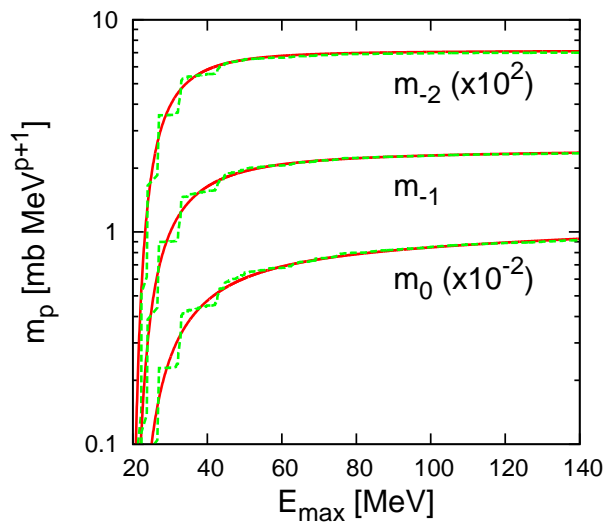,scale=1.3}
\caption{(Color online) Moments of the photoabsorption cross section 
as a function of the upper limit of the integration. 
The moments with the discretized states 
($\theta=0^{\circ}$) are also plotted. The AV8$^\prime$+3NF potential 
is used. Unit of $m_{p}$ is mb\,MeV$^{p+1}$. }
\label{SR.moment.fig}
\end{center}
\end{figure}

Figure~\ref{SR.moment.fig} displays the convergence of the various
moments with respect to the upper limit of the integration. It is 
surprising that even the moments calculated from the discretized states 
with $\theta=0^{\circ}$ lead to a good approximation, already at 
$E_{\rm max}$=60\, MeV, to the moments 
obtained with the CSM. The moment $m_{-2}$ converges well at the rest
energy of a pion, but the moment $m_{0}$ is still 
increasing beyond that energy. Our moments appear
consistent with those calculated with the potential of Argonne
$v$18+UIX in the LIT method~\cite{gazitb}.  
For $E_{\rm max}$=135\,MeV
we obtain $m_{-2}$=0.0710 mb MeV$^{-1}$, $m_{-1}$=
2.36 mb (96\% of $m_{-1}(\infty)$), 
and $m_0$=92.0 mb MeV (73\% of $m_{0}(\infty)$) 
with AV$8^{\prime}$+3NF, while the
corresponding values with G3RS+3NF are 0.0725 mb MeV$^{-1}$, 
2.45 mb (96\% of $m_{-1}(\infty)$), and 
97.1 mb MeV (80\% of $m_{0}(\infty)$), respectively.
We estimate the $m_0$ value for $E_{\rm max}$=135\,MeV 
using the experimental cross sections~\cite{arkatov}. The extracted
value is 100$\pm$5 mb MeV, which agrees fairly well 
with our theoretical values noted above. See also 
Fig.~\ref{photo-decomp.3NN.fig} later.

\subsection{Comparison with experiment}
\label{comparison}

\begin{figure}[th]
\epsfig{file=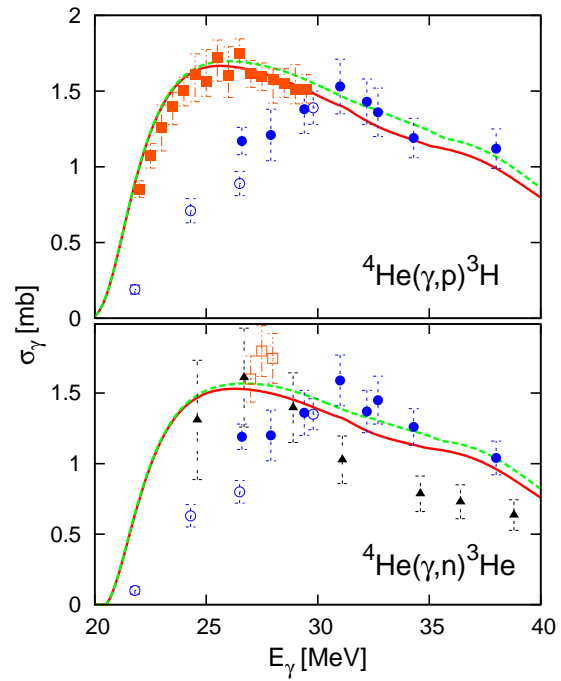,scale=1.2}
\caption{(Color online) Photoabsorption cross sections of  
$^{4}$He$(\gamma,p)$$^3$H and $^4$He$(\gamma,n)$$^3$He 
reactions compared between the MRM calculation and experiment. 
Solid curve: AV8$^\prime$+3NF; dashed curve: G3RS+3NF. 
The data are taken as follows: open circle~\cite{shima}, square~\cite{tornow}, 
closed circle~\cite{shima.new}, and triangle~\cite{nilsson}.}
\label{photo-excl.fig}
\end{figure}

We compare in Fig.~\ref{photo-excl.fig} 
the photoabsorption cross sections for the reactions  
$^{4}$He$(\gamma,p)$$^3$H and  $^4$He$(\gamma,n)$$^3$He between 
the MRM calculation and experiment.   
The calculated cross sections do not depend on the potentials 
up to about 25 MeV, and then  
the AV8$^\prime$+3NF potential predicts slightly smaller values 
than the G3RS+3NF potential beyond the resonance peak. 
The MRM result for the $(\gamma,p)$ cross section 
agrees rather well 
with the 1983 evaluation~\cite{calarco} as well as the very recent
data~\cite{tornow}, but disagrees with the data~\cite{shima} in 
the low energy region. The $^4$He$(\gamma,n)$$^3$He cross 
section obtained in the MRM is much larger than that of 
the 1983 evaluation in the energy 
region of 25--35\,MeV. The $(\gamma, n)$ data are considerably scattered 
among the experiments. Compared to the recent 
data~\cite{nilsson}, the MRM result is consistent with experiment in
25--28\,MeV but is considerably larger than experiment in 30--40\,MeV. 
The MRM cross sections of both $(\gamma,p)$ and $(\gamma,n)$ reactions 
agree fairly well with the new data~\cite{shima.new} beyond 30\,MeV. 
Compared to the LIT calculation~\cite{quaglioni04} with the simple 
Malfliet-Tjon potential, our cross sections of both $(\gamma, p)$ and 
$(\gamma, n)$ are  similar to their results that include final 
state interactions. It is noted that in our MRM calculation 
the peak height of the giant resonance is slightly 
lower and the resonance width is slightly broader than in the LIT 
calculation.

\begin{figure}[th]
\epsfig{file=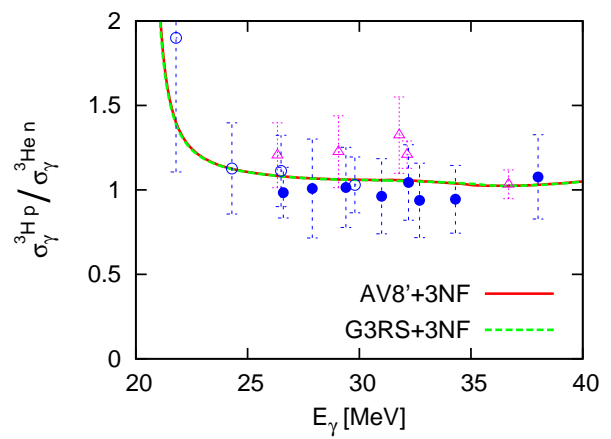,scale=1.3}
\caption{(Color online) Ratio of the photoabsorption cross sections of  
$^{4}$He$(\gamma,p)$$^3$H and $^4$He$(\gamma,n)$$^3$He. The data are 
taken as follows: open circle~\cite{shima}, closed circle~\cite{shima.new}, 
and triangle~\cite{florizone}. }   
\label{photo-ratio.fig}
\end{figure}

Figure~\ref{photo-ratio.fig} shows the cross section ratio of 
$\sigma_{\gamma}^{^3{\rm H}p}/\sigma_{\gamma}^{^3{\rm He}n}$, which 
is an important quantity to test the charge symmetry of the nuclear 
interaction. 
In the present calculation only the Coulomb potential breaks the 
charge symmetry. Both of the AV8$^{\prime}$+3NF 
and G3RS+3NF potentials give 
virtually the same ratio. The calculated ratio is consistent with the recent 
measurements~\cite{shima,shima.new,florizone} as well as the theoretical 
calculations~\cite{unkelbach,quaglioni04}. It is interesting to note that the 
ratio of the data~\cite{shima} agrees very well with our 
result though each of $(\gamma,p)$ and $(\gamma,n)$ cross sections is  
considerably smaller than our cross section.
According to the 1983 evaluation~\cite{calarco} 
the ratio is as large as 1.5 in the 25--30 MeV region. We think it is 
probably attributed to the inefficiency of observing the cross section of 
$^4$He$(\gamma,n)$$^3$He compared to that of $^{4}$He$(\gamma,p)$$^3$H. 
The rise of the ratio below 22 MeV is simply due to the 
difference between the $^3$H+$p$ and $^3$He+$n$ thresholds.

\begin{figure}[th]
\epsfig{file=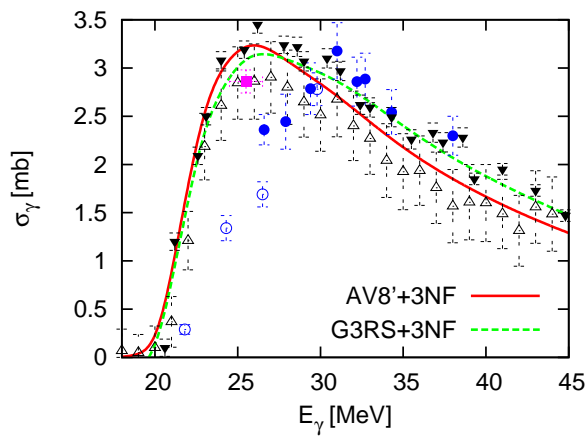,scale=1.3}
\caption{(Color online) Comparison of the photoabsorption cross section 
between the CSM calculation with $\theta$=17$^\circ$ and experiment. 
The data are taken as follows: closed triangle~\cite{arkatov}, 
square~\cite{wells}, open circle~\cite{shima}, 
closed circle~\cite{shima.new}, and open triangle~\cite{nakayama}.}  
\label{photo.fig}
\end{figure}

We display in Fig.~\ref{photo.fig} the total photoabsorption 
cross section 
$\sigma_{\gamma}(E_{\gamma})$ calculated with the CSM. The two potentials 
give qualitatively the same results, but a careful look shows that 
the resonance energy and the width given by the 
AV8$^\prime$+3NF potential are slightly smaller than those 
obtained with the G3RS+3NF potential. 
The calculation predicts a sharp rise of the cross section 
from the threshold, which is observed by several 
measurements~\cite{arkatov,nakayama} but not in the 
data of Ref.~\cite{shima}. 
Our result is consistent with the LIT 
calculations~\cite{gazit,bacca} 
starting from the realistic interactions especially in the cross section
near the threshold. It seems that the data~\cite{nakayama,shima.new} 
indicate a small rise of the cross section at 31\,MeV, but no such behavior 
appears in the theory. 

It is noted that the data of Ref.~\cite{nakayama} appear to predict 
slightly smaller cross sections than our result though the 
shape of the cross section agrees well. This experiment 
is actually not a direct measurement using photons but is 
based on the excitation of the analog states of $^4$He via the 
$^4$He($^7$Li,$^7$Be) reaction. By an ingenious technique to 
separate spin-nonflip cross sections from spin-flip cross sections, 
the $\sigma_{\gamma}(E_{\gamma})$ values were deduced, apart from 
an overall multiplicative factor.  
The factor was fixed by comparing to the sum 
of $(\gamma, p)$ and $(\gamma, n)$ cross sections 
at 40\,MeV that are taken from the 1983 evaluation~\cite{calarco}. 
The factor could be slightly larger, however, 
if it were determined according to the data of 
Refs.~\cite{arkatov,shima.new} and/or if the $\sigma_{\gamma}(E_{\gamma})$ 
value at 40\,MeV were contributed from a 
partial cross section $(\gamma, pn)$~\cite{shima}.  
Then it is probable that the agreement between 
experiment and theory becomes more perfect. 

\begin{figure}[th]
\epsfig{file=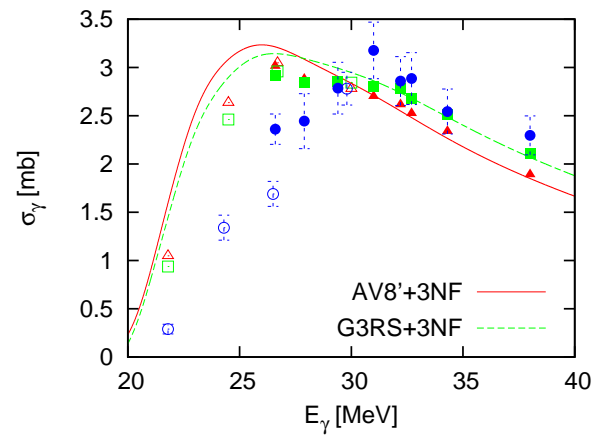,scale=1.3}
\caption{(Color online) The photoabsorption 
cross section weighted by the intensity distributions of photons 
as used in the experiment~\cite{shima,shima.new}. Solid and dotted 
curves are the original theoretical cross sections. 
Open and closed triangles are the weighted 
cross sections with AV8$^{\prime}$+3NF, 
while open and closed squares are the results with G3RS+3NF. Open and
 closed circles are the data taken from Refs.~\cite{shima,shima.new}.} 
\label{Shima.new.exp.fig}
\end{figure}

As noted above, serious disagreement between the theory and the
experiment~\cite{shima} is observed at the energy below 30\,MeV. Other 
theoretical calculations~\cite{quaglioni04,gazit,bacca,quaglioni} 
with the LIT also disagree with the experiment. The 
experiment makes use of pulsed photons produced via the Compton 
backscattering of laser photons with high-energy electrons. The 
cross sections for the two- and three-body decay channels 
were measured in an event-by-event mode. Since the photons have some 
intensity distributions with respect to $E_{\gamma}$, the cross 
section measured is actually a weighted mean of ideal cross 
sections that are free from the spread of the photon energies and 
are to be compared to the theoretical cross section. To see the extent
to which the energy averaging changes the cross section, 
we have calculated such cross sections 
that are weighted by the same distribution functions as used in 
Refs.~\cite{shima,shima.new}. Figure~\ref{Shima.new.exp.fig} 
compares the $\sigma_{\gamma}(E_{\gamma})$ values calculated in this way with 
experiment. The weighting procedure 
gives a different effect on the cross section between below and above 
30\,MeV: Above 30\,MeV the 
original theoretical cross sections change only little 
and very good agreement with experiment is attained 
except for the data at 31\,MeV. In contrast 
to this, below 30\,MeV the cross sections tend to decrease toward 
the experimental data points. However, the decrease is not large 
enough to fill the gap between theory and experiment. The 
disagreement observed at the low energy region still remains to be 
accounted for. The experimental method for the generation of incident
gamma-rays used in 
Refs.~\cite{shima,shima.new} appears to be very similar to that of 
Ref.~\cite{tornow}, but the method of detecting the particles after 
the photoabsorption is different. We hope that the 
discrepancy at the low energy 
region will be resolved experimentally. 

\begin{figure}[th]
\epsfig{file=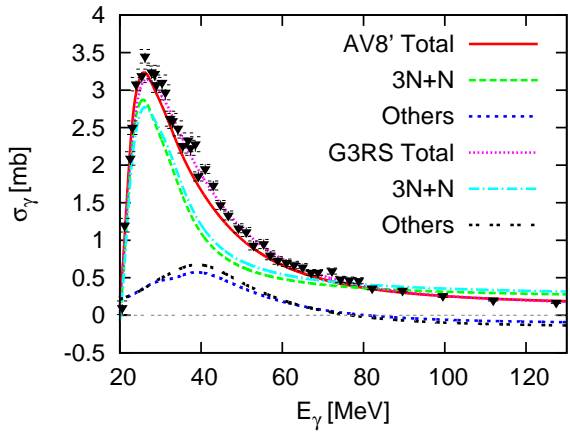,scale=1.3}
\caption{(Color online) Decomposition of the photoabsorption cross 
section into the $3N+N$ and other contributions. See the text for 
how the decomposition is made. The data are taken from
 Ref.~\cite{arkatov}. } 
\label{photo-decomp.3NN.fig}
\end{figure}

It is interesting to compare the cross section at high energy. The
calculation at high energy region is hard in the MRM but is 
not difficult in the CSM. A comparison is made in 
Fig.~\ref{photo-decomp.3NN.fig}, where $E_{\gamma}$ reaches the 
rest energy of a pion. The G3RS+3NF potential appears to  
reproduce the data~\cite{arkatov} more precisely 
between 40--80\,MeV, but in the 
other energy region both potentials give equally good results. 
As mentioned in Sec.~\ref{E1.sum.rule}, the integrated cross 
section $m_0$ for $E_{\rm max}$=135\,MeV is found to agree 
fairly well with the value estimated using the experimental 
cross sections.  
Since the CSM calculation reproduces the total photoabsorption 
cross section, it is meaningful to analyze the contribution of 
the two-body decay channels to $\sigma_{\gamma}(E_{\gamma})$. 
This decomposition 
can be performed by restricting the sum over 
the eigenstates $\lambda$ in Eq.~(\ref{csm.strength}) 
to those whose complex energies lie 
on the rotating continua 
starting from the $^3$H+$p$ and $^3$He+$n$ 
thresholds~\cite{moiseyev,myo,CSM}. The cross section labeled as 
$3N$+$N$ in the figure denotes this partial cross
section. It constitutes a major part of the total 
cross section. The 
cross section labeled Others in the figure is obtained by subtracting 
the $3N$+$N$ cross section from the total cross section. 
As seen in the figure, this quantity is not necessarily positive because
of an interference effect. 
It consists of the three- and four-body decay 
contributions and more importantly of the interference term of 
the two-body and other decay amplitudes. 

\begin{figure}[th]
\epsfig{file=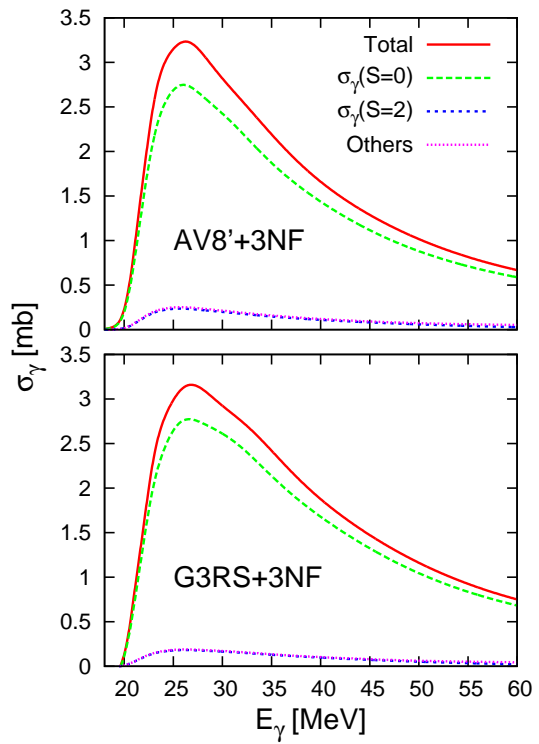,scale=1.2}
\caption{(Color online) Decomposition of the photoabsorption cross 
section into contributions specified by the total spin $S$. See the text
 for detail. }
\label{photo-decomp.SD.fig}
\end{figure}

Another physically interesting decomposition of $\sigma_{\gamma}(E_{\gamma})$ 
is to make use of the total spin $S$. As listed in 
Table~\ref{spectrum3bf}, the ground state of $^{4}$He contains 
by more than 85\% the $S$=$0$ main component and the $S$=$2$ minor component. 
The $S$=$1$ component is negligible. Since the $E1$ operator does 
not change the spin, it makes sense to decompose 
the $\sigma_{\gamma}(E_{\gamma})$ value 
according to the spin channels. Figure~\ref{photo-decomp.SD.fig} displays 
the decomposition into the partial contributions of 
$\sigma_{\gamma}(S$=$0)$, $\sigma_{\gamma}(S$=$2)$, and Others, where 
Others denotes not only the $S$=$1$ contribution but also the 
interfering terms between the different spin amplitudes. 
This partial cross section, e.g. $\sigma_{\gamma}(S$=$0)$, is sensitive
to the probability of finding the $S$=$0$ (and at the same time $L$=$0$) 
components in the ground state of $^4$He. If this partial cross section 
can be measured experimentally, it would give us information on the 
$D$-state probability in the ground state of $^4$He, which is closely 
related to the strength of the tensor force of the $NN$ interaction.

\section{Conclusions}
\label{conclusion.sec}

Motivated by the discrepancy in the low-energy data on the 
photoabsorption cross section of $^4$He, we 
have performed  ${\it ab \ initio}$ calculations for the cross section 
using realistic nuclear forces. 
Our approach takes proper account of the most important ingredients 
for a description of four-nucleon dynamics, e.g., correlated motion 
of the nucleons in both the ground and continuum states of $^4$He, 
effects of the tensor force, 3$N$+$N$-cluster configurations, 
and final state interactions 
in the process of the photodisintegration. 

We have applied two different methods, the complex scaling method 
(CSM) and 
the microscopic $R$-matrix method (MRM), to obtain the cross section. 
The merit of the CSM 
is that one needs no explicit construction of continuum 
states but nevertheless gets the photoabsorption cross
section in a way similar to bound state problems. 
The reliability of our approach is confirmed by observing that 
the two independent methods lead to virtually the same cross section 
in 20--40 MeV region.

In the energy region between 30--40 MeV
the calculated cross sections for $^4$He($\gamma$, $p$)$^3$H 
and $^4$He($\gamma$, $n$)$^3$He are found to agree with 
the very recent measurements~\cite{tornow,shima.new}. 
The total photoabsorption cross section calculated up to 
the rest energy of a pion is also in fair agreement 
with most of the available data~\cite{arkatov,nakayama,wells,shima.new} 
but the one~\cite{shima,shima.new} in the low energy region of 
20--30 MeV. The calculated total cross section 
sharply rises from the $^3$H+$p$ threshold and reaches a peak at 
about 26--27\,MeV consistently with the Lorentz integral transform 
calculations, but in disagreement with the data~\cite{shima,shima.new}.  
Hoping to resolve this discrepancy, we have allowed for 
the energy spread of 
the photon beams in the measurement and calculated 
the energy-averaged cross sections using the 
the same distribution functions as those in Ref.~\cite{shima,shima.new}. 
The cross sections in fact decrease below 30\,MeV but it turns out 
that the change is not large enough to account for the discrepancy. 

The configurations that the MRM calculation takes in the internal region
are represented by several two-cluster partitions. The 
$^3$H+$p$ and $^3$He+$n$ cluster configurations, among others, 
are most important to reproduce the photoabsorption cross section 
in the energy region of 20--35 MeV. This is further corroborated from the 
analysis of the transition densities as well as the decomposition 
of the cross section into the $3N$+$N$ contribution. 

The electric dipole transition occurs mainly from 
the major component with $(L, S)$=(0, 0) of the $^4$He ground state to 
the (1, 0) component of the $1^-1$ continuum states. It should be 
noted, however, that both the ground and $1^-1$ excited states 
gain energy largely from the tensor force, and in fact we 
have seen an important role of the tensor force induced by the one-pion
exchange 
in enhancing the photoabsorption cross section as well as 
the photonuclear sum rule. 

In this work we have presented the analysis of the 
electric dipole strength function.
A similar analysis for $^4$He 
will be possible for other strength functions induced by, e.g.,   
Gamow-Teller and spin-dipole operators that 
probe different spin-isospin responses of $^4$He. 
A study along this direction is underway and will be reported elsewhere.

\section*{Acknowledgments}

We thank T. Shima for many valuable communications and for 
making some new data available to us prior to publication. 
Thanks are also due to S. Nakayama and S. Aoyama for useful 
discussions. 
W.H. is supported by the Special Postdoctoral Researchers Program of
RIKEN. The work of Y.S. is supported in part by a Grant-in-Aid for 
Scientific Research (No. 21540261) of Japan Society for the 
Promotion of Science.

\end{document}